\documentclass{PoS}

\def\be{\begin{equation}}
\def\ee{\end{equation}}
\def\ba{\begin{eqnarray}}
\def\ea{\end{eqnarray}}

\def\dalemb#1#2{{\vbox{\hrule height.#2pt
        \hbox{\vrule width.#2pt height#1pt \kern#1pt \vrule width.#2pt}
        \hrule height.#2pt}}}

\def\dalemb#1#2{{\vbox{\hrule height.#2pt
        \hbox{\vrule width.#2pt height#1pt \kern#1pt \vrule width.#2pt}
        \hrule height.#2pt}}}

\def\ba{\begin{eqnarray}}
\def\ea{\end{eqnarray}}
\def\be{\begin{equation}}
\def\ee{\end{equation}}

\def\gtorder{\mathrel{\raise.3ex\hbox{$>$}\mkern-14mu
             \lower0.6ex\hbox{$\sim$}}}
\def\ltorder{\mathrel{\raise.3ex\hbox{$<$}\mkern-14mu
             \lower0.6ex\hbox{$\sim$}}}

\title{Non-minimal coupling as a mechanism for spontaneous symmetry breaking
  on the brane}

\ShortTitle{Non-minimal coupling as a mechanism for spontaneous symmetry
  breaking}

\author{\speaker{Orfeu Bertolami}${^a}$\thanks{Also at
Centro de F\'\i sica dos Plasmas, Instituto Superior T\'ecnico.}
and Carla Carvalho${^{ab}}$\thanks{Also at
Centro de F\'\i sica dos Plasmas, Instituto Superior T\'ecnico.}\\
\llap{$^a$}Departamento de F\'\i sica, Instituto Superior T\'ecnico \\
Avenida Rovisco Pais 1, 1049-001 Lisboa, Portugal\\
\llap{$^b$}Laboratoire Physique Th\'eorique,\\
B\^atiment 210, 91405 Orsay, France\\
E-mail: \email{orfeu@cosmos.ist.utl.pt}, \email{ccarvalho@ist.edu}}

\abstract{Motivated by the dimensional asymmetry characteristic of
  braneworlds, we populate the bulk spacetime with matter scalar fields, both
  real and complex, and couple them non-minimally to gravity. We derive the
  effective equations of motion on the brane and realize the case when the
  fields acquire a non-vanishing vacuum expectation value. This entails a
  change in the effective cosmological constant and in the effective mass of
  the scalar fields. We find that the non-minimal coupling provides a
  mechanism for generating spontaneous symmetry breaking at Planck energies
  on the brane. This mechanism is however rendered unviable for the
  generation of masses at the energy scales of the standard model.}

\FullConference{From Quantum to Emergent Gravity: Theory and Phenomenology\\
        June 11-15 2007\\
        Trieste, Italy}

\begin{document}


\section{Introduction}

Braneworld cosmological models, by realizing the four-dimensional
universe entirely in a hypersurface of a higher dimensional space, are
intrinsically dimensionally asymmetric.  The emergence of the
difference between the parallel and the orthogonal directions to this
hypersurface seems to reverberate the Goldstone mechanism, where
massless multiplets are imparted with a mass in the direction along
which symmetry is spontaneously broken. Here we propose to use
spontaneous symmetry breaking (SSB) of bulk scalar fields to relate
the mass measured on the brane to a bulk mechanism and thus look for
signatures of extra dimensions. This is the opposite idea to the
Dvali-Shifman mechanism, where a non-confinement Higgs (spontaneously
broken symmetry) phase is found on the brane, whereas a confinement
(restored symmetry) phase is preserved in the bulk \cite{shifman97}.
Moreover, we couple the bulk scalar fields non-minimally to the Ricci
scalar. This is the simplest interaction which yields a canonical
kinetic term. Furthermore, it reduces to Brans-Dicke up to a field
transformation for a vanishing vacuum expectation value (vev).
Recently, a variation of the Dvali-Shifman mechanism triggered by the
geometry of five-dimensional anti-de Sitter was applied to the study
of the dynamics along the direction normal to the brane of a scalar
field non-minimally coupled to gravity \cite{farakos05}.

This contribution reports on a recent study where we consider both
real and complex scalar fields in the bulk space and examine their
implication in the mechanism of SSB on the brane upon acquiring a
non-vanishing bulk vev \cite{BC07}. Furthermore, when studying the
case of the complex scalar field, we also consider a minimally coupled
$U(1)$ gauge field so that spontaneous breaking of the gauge symmetry
can take place. After establishing how bulk quantities induce
quantities on the brane, we compute the equations of motion induced on
the brane starting from a bulk action. The resulting Einstein
equations provide the relations between the induced geometry on the
brane and the matter fields therein.  Furthermore, we observe that
matter {\it a priori} localized on the brane, such as the brane
tension, will only interact gravitationally with the bulk matter
fields induced on the brane when a non-minimal coupling
exists. Otherwise, brane and bulk matter fields do not see each
other. The non-minimal coupling also reinforces the canonical
mechanism of spontaneous symmetry breaking at very high energies via
the interaction of the brane tension with the scalar fields. Moreover,
we discuss whether the SSB mechanism could suffice to generate masses
on the brane and thus provide a localization mechanism. The
four-dimensional masses, both of the scalar and the gauge fields,
induced on the brane are found to be of order the four-dimensional
Planck mass. Taking the inverse of the mass as a measure of the
confinement to the brane \cite{shaposhnikov04}, it follows that the
range of the induced interaction is short about the brane, which
suggests the localization of the bulk fields about the position of the brane.

In this contribution we consider a real scalar field in the bulk and
non-minimally coupled to gravity via the Ricci scalar. 
[For the case of the complex scalar field, we refer the reader to
Ref.~\cite{BC07}.] 
Following the procedure described in the Appendix, we derive the
induced equations on the brane from the bulk action.
In particular, we derive the effective potential on the brane and
compute the effective mass of the scalar field induced on the brane
upon taking a non-vanishing vev in the bulk. 
Finally, we analyse the implications of our results for a mechanism of
localization of matter on the brane. 
We keep the number of space dimensions $d$ arbitrary, rendering the
results valid for any codimension-one brane.

\section{Real Scalar Field Non-minimally Coupled in the Bulk}
\label{sec:scalar}

In this section we study the case of a bulk, real scalar field $\phi.$
In addition to the Einstein-Hilbert term and the canonical kinetic and
potential terms of $\phi,$ we consider an interaction term with a
non-minimal coupling of $\phi$ to the Ricci scalar 
\ba {\cal L}
={1\over {\kappa_{(5)}^2}}R -2\Lambda +\xi\phi^2R -{1\over
2}g^{\mu\nu}(\nabla_{\mu}\phi)(\nabla_{\nu}\phi) -V(\phi^2).
\label{eqn:lagrangian:Phi}
\ea  
Here, $\kappa_{(5)}^2=8\pi G_{N(5)}=1/M_{Pl}^3$ is the
five-dimensional gravitational coupling constant and $\xi$ is a
dimensionless coupling constant which measures the non-minimal
interaction. In the cosmological constant term $\Lambda =\Lambda_{(5)}
+\Lambda_{(4)}$ we have included both the bulk vacuum value
$\Lambda_{(5)}$ and that of the brane $\Lambda_{(4)},$ described by a
brane tension $\sigma$ localized at the position of the brane,
$\Lambda_{(4)}=\sigma\delta(N).$

\subsection{The induced dynamics on the brane}

First we derive the equations of motion for both the scalar field and the
graviton as measured by an observer localized on and confined to the
brane. This procedure follows closely the techniques developed in
Ref.~\cite{BC06} for a vector field and in particular uses the results derived
therein, which for completion are included in an appendix.

By varying the action with respect to the metric, we obtain the
Einstein equation in the bulk
\ba
\left( {1\over \kappa_{(5)}^2} +\xi\phi^2\right)G_{\mu\nu}
+\Lambda g_{\mu\nu}
={1\over 2}T^{(\phi)}_{\mu\nu}
+\xi\Sigma^{(\phi)}_{\mu\nu} ~,
\ea
where
\ba
T^{(\phi)}_{\mu\nu}
=(\nabla_{\mu}\phi)(\nabla_{\nu}\phi)
+g_{\mu\nu}\left[
-{1\over 2}g^{\alpha\beta}(\nabla_{\alpha}\phi)(\nabla_{\beta}\phi)
-V(\phi^2)\right]
\ea
is the stress-energy tensor associated with $\phi$ and
\ba
\Sigma^{(\phi)}_{\mu\nu}
=\nabla_{\mu}\nabla_{\nu}\phi^2
-g_{\mu\nu}g^{\alpha\beta}\nabla_{\alpha}\nabla_{\beta}\phi^2
\ea
is the contribution from the interaction term.
We note the
$\phi$--dependence of the five-dimensional gravitational coupling
constant akin to that of the Brans-Dicke formulation.
For the equation of motion for the $\phi$ field, obtained by
varying the action with respect to $\phi,$ we find that
\ba
g^{\mu\nu}\nabla_{\mu}\nabla_{\nu}\phi
-{\partial V\over \partial \phi}
+2\xi\phi R
=0~.
\ea

We now proceed to project the equations in the directions parallel
(denoted by $A$) and orthogonal (denoted by $N$) to the surface of the
brane, finding for the stress-energy tensor ${\bf T ^{(\phi)}}$ that  
\ba
T_{AB}^{(\phi)}
&=& (\nabla_{A}\phi)(\nabla_{B}\phi) +g_{AB}\left[ -{1\over 2}\left[
(\nabla_{C}\phi)^2 +(\nabla_{N}\phi)^2\right] -V(\phi^2)\right],\cr 
T_{AN}^{(\phi)}
&=& (\nabla_{A}\phi)(\nabla_{N}\phi)~, \cr 
T_{NN}^{(\phi)}
&=& (\nabla_{N}\phi)(\nabla_{N}\phi)
+g_{NN}\left[ -{1\over 2}\left[ (\nabla_{C}\phi)^2 +(\nabla_{N}\phi)^2\right]
-V(\phi^2)\right], 
\ea 
and similarly for the source tensor $\Sigma ^{(\phi)}_{\mu \nu}$
that
\ba
\Sigma^{(\phi)}_{AB} &=&\left( \nabla_{A}\nabla_{B}
+K_{AB}\nabla_{N}\right)\phi^2 -g_{AB}\left( \nabla_{C}^2 +\nabla_{N}^2
+K\nabla_{N}\right)\phi^2~, \cr 
\Sigma^{(\phi)}_{AN} &=&\left( \nabla_{A}\nabla_{N}
-K_{A}{}^{B}\nabla_{B}\right)\phi^2~,\cr 
\Sigma^{(\phi)}_{NN} &=&\nabla_{N}\nabla_{N}\phi^2
-g_{NN}\left( \nabla_{C}^2 +\nabla_{N}^2 +K\nabla_{N}\right)\phi^2~. 
\ea 
Equating the $(AB)$ components of the decomposition of the Einstein
tensor and of the source terms from 
the scalar field $\phi,$ we find for the Einstein equation parallel
projected on to the brane that 
\ba 
&&\left( {1\over \kappa_{(5)}^2} +\xi \phi^2\right)
 \left[ G_{AB}^{(ind)} +2K_{AC}K_{B}{}^{C} -K_{AB}K -K_{AB,N}
-{1\over 2}g_{AB}\left(-K_{CD}K^{CD} -K^2 -2K_{,N}\right)\right]
\cr
&=&{1\over 2}(\nabla_{A}\phi)(\nabla_{B}\phi)
+{1\over 2}g_{AB}\left[
-{1\over 2}\left[ (\nabla_{C}\phi)^2 +(\nabla_{N}\phi)^2\right]
-V(\phi^2) \right] - g_{AB}\Lambda\cr
&+&\xi\left[
\left( \nabla_{A}\nabla_{B} +K_{AB}\nabla_{N}\right)\phi^2
-g_{AB}\left( \nabla_{C}^2 +\nabla_{N}^2
+K\nabla_{N}\right)\phi^2\right].
\label{eqn:einsteinAB:Phi}
\ea

To obtain the matching condition for the extrinsic curvature across the
brane, we integrate the $(AB)$ component of the Einstein equation in the
coordinate normal to the brane.
For a $Z_{2}$--symmetric brane, we obtain for the $(AB)$ matching
condition across the brane
\ba
&&\int _{-\delta}^{+\delta}dN
\left( {1\over \kappa_{(5)}^2} +\xi \phi ^2\right)
 \left( -K_{AB,N} +g_{AB}K_{,N}\right)\cr
&=&\int _{-\delta}^{+\delta} dN\biggl[
-g_{AB}\Lambda_{(4)}
+\xi \left(
\left( K_{AB} -g_{AB}K\right)\nabla_{N}\phi ^2
-g_{AB}\nabla_{N}^2\phi ^2 \right)\biggr],
\ea
which yields
\ba
\left( {1\over \kappa_{(5)}^2} +\xi \phi ^2\right)
 \left( -K_{AB} +g_{AB}K\right)
= g_{AB}\left( -{\sigma\over 2} -\xi \nabla_{N}\phi ^2\right)
\label{eqn:imc:Phi}
\ea
for $\phi ^2$ even about the position of the brane.
These provide boundary conditions for ten of the fifteen degrees of
freedom. Five additional boundary conditions are provided by the
matching conditions from the $(AN)$ and $(NN)$ components of the projected
Einstein equations.
From inspection of the $(AN)$ components we note that on the brane
\ba
G_{AN}
=K_{A}{}^{B}{}_{;B} -K_{;A}
=-\nabla_{B}\left( \int dN~G_{A}{}^{B}\right) 
=-\nabla_{B}{\cal T}_{A}{}^{B (\phi)} =0~,
\ea
which must vanish in order to preserve conservation of the induced
stress-energy 
${\cal T}_{AB}^{(\phi)}$ on the brane, as read off of the right-hand
side of Eq.~(\ref{eqn:imc:Phi}).
This condition constrains four degrees of freedom.
The $(NN)$ component of the Einstein equation
\ba
&&\left( {1\over \kappa_{(5)}^2} +\xi \phi ^2\right)\left[
-R^{(ind)} -K_{CD}K^{CD} +K^2\right] +\Lambda _{(5)}\cr
&=&{1\over 2}\left[
{1\over 2}\left( \nabla_{N}\phi\right)\left( \nabla_{N}\phi\right)
-{1\over 2}\left( \nabla_{C}\phi\right)\left( \nabla_{C}\phi\right)
-V(\phi^2)\right]
-\xi\left( \nabla_{C}^2 +K\nabla_{N}\right)\phi ^2
\label{eqn:einsteinNN:Phi}
\ea
consists of the remaining constraint.

Similarly, we expand the equation of motion for the scalar field
$\phi$
\ba
&&\left[ g^{AB}\left( \nabla_{A}\nabla_{B} +K_{AB}\nabla_{N}\right)
+\nabla_{N}\nabla_{N} \right]\phi
-{\partial V\over \partial \phi}\cr
&+&2\xi\phi
 \left( R^{(ind)} -K_{AB}K^{AB} -K^2 -2K_{,N}\right)
=0 ~.
\label{eqn:Phi}
\ea
To obtain the matching condition for $\phi$ across the brane,
we integrate in the $N$ coordinate discarding all derivatives other
than along $N$
\ba
&&\int _{-\delta}^{+\delta}dN\left[
K\nabla_{N}\phi
+\nabla_{N}\nabla_{N}\phi
-4\xi\phi K_{,N}\right] =0~.
\ea
If we assume $Z_{2}$-symmetry across the brane,
the matching condition for $\phi$ becomes
\ba
\nabla_{N}\phi -4\xi K\phi =0 ~.
\label{eqn:Phi:mc}
\ea
Substituting Eq.~(\ref{eqn:Phi:mc}) back into Eq.~(\ref{eqn:Phi}),
we obtain for the propagation of $\phi$ on the brane that
\ba
g^{AB}\nabla_{A}\nabla_{B}\phi
-{\partial V\over \partial \phi}
+2\xi\phi
 \left[ R^{(ind)} -K_{AB}K^{AB} +\left( 1 +8\xi\right)K^2\right]
=0~.
\label{eqn:Phi:ind}
\ea

Moreover, equating the matching conditions for the extrinsic curvature,
Eq.~(\ref{eqn:imc:Phi}), and for the scalar field $\phi,$
Eq.~(\ref{eqn:Phi:mc}), we can solve for $K_{AB}$ and $\nabla_{N}\phi$
to find that
\ba
K_{AB}
&=&-{1\over 2}g_{AB}~\sigma
{1/(d -1)\over
 {1/\kappa _{(5)} ^2 +\xi \phi ^2 [1 +8\xi d/(d-1)]}}\vert_{N=0}~,\\
\nabla_{N}\phi\vert_{N=0}
&=&-2\xi\phi~\sigma
{d/(d -1)\over
 {1/\kappa _{(5)} ^2 +\xi \phi ^2[ 1 +8\xi d/(d -1)]}}\vert_{N=0}~.
\ea
We substitute Eq.~(\ref{eqn:Phi:mc}) for $\nabla_{N}\phi$ and
Eq.~(\ref{eqn:Phi:ind}) for $\nabla_{C} ^2\phi$ in the $(NN)$ component
of the Einstein
equation, Eq.~(\ref{eqn:einsteinNN:Phi}), to find for $R^{(ind)}$ that
\ba
&&\left(
{1\over \kappa_{(5)}^2}
+\xi\phi^2\left( 1 +4\xi\right)\right)R^{(ind)}
=\left( {1\over 4} +2\xi\right)(\nabla_{C} \phi)^2
+{1\over 2}V
+2\xi\phi{\partial V\over \partial \phi} +\Lambda_{(5)} \cr
&-&K_{CD}K^{CD}\left(
{1\over \kappa_{(5)}^2}
+\xi\phi^2\left( 1 -4\xi\right)\right)
+K^2\left(
{1\over \kappa_{(5)}^2}
+\xi\phi^2\left( 1 -32\xi ^2\right)\right)~.
\label{eqn:Rind:Phi}
\ea
Similarly, from Eq.~(\ref{eqn:einsteinAB:Phi}) we find
for the Einstein equation induced on the brane that
\ba
G_{AB}^{(ind)}
&=&\left( {1\over \kappa_{(5)}^2} +\xi \phi^2\right)^{-1}
 \left[\left( {1\over 2} +2\xi\right)(\nabla_{A}\phi)(\nabla_{B}\phi)
 +2\xi \phi\nabla_{A}\nabla_{B}\phi\right]\cr
&-&g_{AB}
\left[ R^{(ind)} +K^2{ {-d^2 +d +4}\over {2d^2}} \right]~,
\label{eqn:Einstein-phi}
\ea
with $R^{(ind)}$ given by Eq.~(\ref{eqn:Rind:Phi}).

Using the equations derived above, we realize the case where the scalar
field acquires a non-vanishing vev
$\left< \phi \right>$ which minimizes the
effective potential. This can induce spontaneous symmetry breaking
when the scalar field is coupled to a gauge field, thus endowing the
latter with a mass,
as discussed in Ref. \cite{BC07}.
Here, however, a non-vanishing vev will entail a change in the
effective cosmological constant
and in the effective mass of the scalar field.
Moreover, once the scalar field acquires a vev, no
direction on the brane can be selected, which implies that
$\nabla_{A}\left< \phi \right> =0.$ Consequently, Lorentz symmetry breaking
cannot take place in the presence of a bulk scalar field only. [See
Ref. \cite{BC06} for
the case of an explicit violation of Lorentz symmetry due to a
non-vanishing vev for a vector field.]

We can read off of the induced Einstein equation the effective cosmological
constant, which would
comprise all the terms proportional to the induced metric which do
not vanish when all the contributions from the matter fields
vanish.
However, in the case that the matter
fields acquire a non-vanishing vev, the effective
cosmological constant will contain the contribution of the matter
fields at the corresponding non-vanishing value. It follows that
\ba
&&\Lambda _{eff}\left(
{1\over\kappa _{(5)}^2} +\xi \left<\phi\right>^2(1 +4\xi)\right)
=
{1\over 2}V(\left<\phi\right> ^2)
+2\xi \left<\phi\right> {\partial V\over \partial \phi}
 \Bigg|_{\phi =\left<\phi\right>}
+\Lambda _{(5)}\cr
&+&K^2|_{\phi =\left<\phi\right>}
\left[
\left( {1\over \kappa_{(5)}^2} +\xi \left<\phi\right>^2\right)
 { { d^2 -d +4}\over {2d^2}}
+(2\xi \left<\phi\right>)^2\left(
{ { -d^2 +3d +4}\over {2d^2}} -8\xi\right)
\right].
\label{eqn:Lambda_eff}
\ea
We thus observe that a non-vanishing vev in the bulk generates in the
gravitational sector a contribution to the cosmological constant on the
brane.

\subsection{The effective potential on the brane}

Whether a non-vanishing vev for the scalar
field can be observed on the brane
depends on the form of the effective potential $V_{eff}(\phi ^2)$.
The parameters of the potential will influence the magnitude of its minimum
and consequently the mass of the scalar field $\phi$ measured on the brane,
defined as the value of the second derivative of the
effective potential evaluated at the vev of the
scalar field, $\left< \phi \right>$.
We first determine the effective potential measured on the brane and
then proceed to study the conditions for a non-vanishing vev.

The evolution of $\phi$ on the brane, as described by
Eq.~(\ref{eqn:Phi:ind}) and with $R^{(ind)}$ given by
Eq.~(\ref{eqn:Rind:Phi}),
\ba
g^{AB}\nabla_{A}\nabla_{B}\phi
-{\partial V_{eff}\over \partial \phi}
+\left( {1\over 4} +2\xi\right)\left( \nabla_{C}\phi\right)^2
 {{2\xi \phi}\over {1/\kappa_{(5)} ^2 +\xi\phi ^2(1 +4\xi)}}
=0
\ea
is determined by a damping term as well as by the effective potential
induced on the brane.
Here,
\ba
-{\partial V_{eff}\over \partial \phi}
&=&{1\over {1/\kappa_{(5)} ^2 +\xi\phi ^2(1 +4\xi)}}
\Biggl\{
-{\partial V\over \partial \phi}
 \left( {1\over \kappa_{(5)} ^2} +\xi\phi ^2\right) \cr
&+&
2\xi\phi \left[
{1\over 2}V
+\Lambda_{(5)}\right]\cr
&+&
2\xi\phi K^2\left[
{1\over \kappa_{(5)} ^2}\left( -{2\over d} +2 +8\xi\right)
+\xi \phi ^2\left( -{2\over d} +2 +12\xi\right)\right]
\Biggr\} ~,\qquad
\label{eqn:dVeff}
\ea
where $V(\phi ^2)$ is the bulk potential, which is assumed to
have a Higgs type form
$V(\phi) =\mu _{(5)} ^2 (\phi^2/2) +\lambda _{(5)} (\phi^4/4)$ with
$\lambda_{(5)} >0$.
We compute $V_{eff}$ by integrating Eq.~(\ref{eqn:dVeff})
to find that
\ba
V_{eff}(\phi ^2)
&=&
\phi  ^2\left[
{\mu _{(5)}^2\over 2}{1\over {1 +4\xi}}
-{\mu _{(5)}^2\over 4} {1\over {1 +4\xi}}
+\lambda _{(5)}{1 \over {\xi \kappa _{(5)}^2}}{1\over {2( 1 +4\xi)^2}}
 \left( {1\over 4} +4\xi\right)\right]\cr
&+&\phi ^4{\lambda _{(5)}\over 4}\left[
{1\over {1 +4\xi}}
-{1\over {4(1 +4\xi)}}\right]
\cr
&+&
\ln\left[ 1  +\xi\kappa _{(5)}^2\phi ^2(1 +4\xi)\right]\times\cr
&&\times\Biggl[
\mu _{(5)}^2{1 \over {\xi \kappa _{(5)}^2}}{1\over {2 (1 +4\xi)^2}}
 \left( {1\over 2} +4\xi\right)
-\lambda _{(5)}{1 \over {(\xi\kappa _{(5)}^2)^2}}
  {1\over {2 (1 +4\xi)^3}}
   \left( {1\over 4} +4\xi\right)\cr
&&
-{1\over {1 +4\xi}}
\Lambda _{(5)}
\Biggr]\cr
&+&{\sigma ^2\over 16}{d^2\over {(d +1)(d -1)}}\times\cr
&&\times\Biggl[
{{[ -2/d +2(1 +4\xi)](8d/(d-1)) -4}\over {1 +8\xi d/(d -1)}}
  {1\over { 1/\kappa _{(5)}^2 +\xi\phi ^2[1 +8\xi d/(d -1)]}}\cr
&&-{\kappa _{(5)}^2\over \xi}{{d -1}\over {d +1}}\left(
  -{2\over d} + 1 +8\xi\right)
   \ln\left[
    {{ 1 +\xi\kappa _{(5)}^2\phi ^2(1 +4\xi)}\over
     {{ 1 +\xi\kappa _{(5)}^2\phi ^2[1 +8\xi d/(d -1)]}}}\right]\Biggr]
\ea
for $\xi \not=-1/4.$
Notice that, in the limit when $\xi \to 0,$ one recovers the original
bulk potential with an extra term on the brane tension,
$V(\phi ^2) +(3/4)\sigma ^2\kappa _{(5)}^2d^2/[(d +1)(d -1)].$
It is natural to expect that there exists a hierarchy of scales
depending on whether
the vev $\left< \phi\right>$ is related to the
Standard Model (SM) scale or the grand unified theory scale. Thus, for
$\vert \xi \phi ^2\vert \ll 1/\kappa_{(5)}^2,$
we can expand the denominator of the first term in $\sigma$
about $1/\kappa_{(5)}^2,$ keeping terms up to order six in $\phi,$
$(\xi \phi ^2)^3.$
The effective potential can thus be written as
\ba
V_{eff}(\phi ^2)
&=& \phi ^2\Biggl\{
{\mu _{(5)}^2\over 2}{1\over {1 +4\xi}}
-{\mu _{(5)}^2\over 4} {1\over {1 +4\xi}}
+\lambda _{(5)} {1 \over {\xi \kappa _{(5)}^2}}{1\over {2( 1 +4\xi)}}
 \left( {1\over 4} +4\xi\right)\cr
&&-{1\over 16}\sigma ^2\kappa _{(5)}^2
  (\xi\kappa _{(5)}^2)
   {d ^2\over {(d +1)(d -1)}}\left[
\left( -{2\over d} +2\left( 1 +4\xi\right)\right)
 {{8d}\over {d -1}} -4\right]
\Biggr\}\cr
&+&\phi ^4 \Biggl\{
{\lambda _{(5)}\over 4}{1\over {1 +4\xi}}
-{\lambda _{(5)}\over 4}{1\over {4(1 +4\xi)}}\cr
&&+{1\over 16}\sigma ^2\kappa _{(5)}^2
  {1\over 2}(\xi\kappa _{(5)}^2)^2
   {d ^2\over {(d +1)(d -1)}}\left[
\left( -{2\over d} +2\left( 1 + 4\xi\right)\right)
{{8d}\over {d -1}} -4\right]
\left( 1 +\xi{ {8d}\over {d -1}}\right)
\Biggr\}\cr
&-&O[\phi ^6]{1\over 16}\sigma ^2\kappa _{(5)}^2{1\over 6}
({\xi\kappa _{(5)}^2})^3\cr
&+&
\ln\left[ 1 +\xi\kappa _{(5)}^2\phi ^2(1 +4\xi)\right]\times\cr
&&\times\Biggl[
\mu _{(5)} ^2{1 \over {\xi \kappa _{(5)}^2}}{1\over {2 (1 +4\xi)^2}}
 \left({1\over 2} +4\xi\right)
-\lambda _{(5)}{1 \over {(\xi\kappa _{(5)}^2})^2}
  {1\over {2 (1 +4\xi)^3}}
   \left( {1\over 4} +4\xi\right)\cr
&&-{1\over {1 +4\xi}}
\Lambda _{(5)}
\Biggr]\cr
&+&{1\over 16}\sigma ^2\kappa _{(5)}^2{d^2\over {(d +1)(d -1)}}\times\cr
&&\times\Biggl[
{{[ -2/d +2(1 +4\xi)](8d/(d-1)) -4}\over {1 +8\xi d/(d -1)}}\cr
&&-{1\over \xi}{{d -1}\over {d +1}}\left(
  -{2\over d} + 1 +8\xi\right)
   \ln\left[
    { { 1 +\xi\kappa _{(5)}^2\phi ^2(1 +4\xi)}\over
     { { 1 +\xi\kappa _{(5)}^2\phi ^2[1 +8\xi d/(d -1)]}}}\right]\Biggr].
\ea
We thus observe that, up to sub-dominant logarithmic terms, the
effective potential is of the form
$V_{eff} ={\mu _{eff} ^2}(\phi ^2/2) +\lambda_{eff} (\phi ^4/4)+
O[\phi ^6],$ where
\ba
\mu _{eff} ^2 &\sim&
\mu _{(5)} ^2
+\lambda_{(5)} {1 \over {\xi \kappa _{(5)}^2}}
-\sigma ^2 \xi \kappa _{(5)}^4~,
\label{eqn:mu_eff}\\
\lambda _{eff} &\sim&
\lambda _{(5)}
+\sigma^2 \xi ^2\kappa _{(5)}^6 ~.
\label{eqn:lambda_eff}
\ea
If $\mu _{eff}^2 <0$ and $\lambda _{eff}>0$, then one expects a
non-vanishing vev for the
scalar field. The first condition guarantees that a non-vanishing
minimum exists,
whereas the second condition guarantees that such minimum is finite.
Conversely, if $\mu _{eff}^2 >0$ and $\lambda _{eff}>0,$ then symmetry
is always unbroken.
Thus, imposing that $\lambda_{eff} >0,$ it follows that
$\lambda _{(5)} > - \sigma^2 \xi ^2\kappa _{(5)}^6.$
Consequently, in order to verify the condition $\mu _{eff}^2 <0,$
we must have that
$\mu _{(5)}^2 < -2\sigma ^2 \xi \kappa _{(5)}^4.$

We notice that the bulk scalar field $\phi,$ being
a five-dimensional field, has dimension
$[\phi]=M^{3/2}$. Accordingly, $\mu _{(5)}$ has dimension of mass and
$\lambda _{(5)}$ dimensions of inverse of mass.
In order to recover characteristically four-dimensional quantities, we
define the four-dimensional scalar field $\Phi$ as the rescaling of
$\phi$ by an appropriate mass scale $M_{\phi}$.
In the mode expansion of a bulk field, this mass can be identified
with the mode function dependent on the direction $N$ evaluated at the
position of the brane in the bulk.
Thus, for $\phi =M_{\phi}^{{1/2}}\Phi,$
the induced equation of motion for $\Phi$ on the brane becomes
\ba
g^{AB}\nabla_{A}\nabla_{B}\Phi
-{1\over M_{\phi}}{\partial V_{eff}\over \partial \Phi}
+ \left( {1\over 4} +2\xi\right)\left( \nabla_{C}\Phi\right)^2
 {{2\xi M_{\phi} \Phi}\over
  {1/\kappa_{(5)} ^2 +\xi M_{\phi}\Phi ^2(1 +4\xi)}}
=0 ~.
\ea
Consequently, the parameters of the effective potential will scale as
\ba
{1\over M _{\phi}}V_{eff}(\Phi ^2)
=\mu _{eff} ^2 {\Phi ^2\over 2} +\lambda _{eff}M _{\phi}{\Phi ^4\over 4}
+M _{\phi} ^2 O[\Phi ^6]~.
\ea
with equations (\ref{eqn:mu_eff}) and (\ref{eqn:lambda_eff}) becoming
\ba
\mu _{eff} ^2 &\sim&
\mu ^2
-2\sigma ^2 \xi \kappa _{(5)}^4~,
\label{eqn:mu:eff2} \\
M_{\phi}\lambda _{eff} &\sim&
\lambda
+M_{\phi}\sigma^2 \xi^2 \kappa _{(5)}^6 ~,
\label{eqn:lambda:eff2}
\ea
where $\mu =\mu _{(5)}$ and $\lambda =M_{\phi}\lambda _{(5)}.$
Here, for $\xi >0$ we have two possible mechanisms for the generation of
a non-vanishing vev: the canonical way, via the potential
associated with the scalar field, and the braneworld way, via the
interaction of the scalar field with the brane tension.
For the latter to be viable in the context of the SM, then
\ba
\left|{\mu _{eff} ^2\over {M_{\phi}\lambda _{eff} }}\right|
\sim {1\over {\xi M_{\phi}}} {1\over \kappa _{(5)}^2}
\ea
must be of order $TeV ^2$, and $\left< \Phi\right> = 246$ GeV.
However, in order to recover the four-dimensional gravitational
coupling constant in Eq.~(\ref{eqn:Einstein-phi}), we find from the
$\phi$ contribution that
$M_{Pl(4)} ^{-2} =\kappa _{(5)}^2M_{\phi},$
and hence that
\ba
M_{Pl} ^3\equiv {1\over \kappa _{(5)}^2}
=M_{Pl(4)} ^2 M_{\phi}.
\label{eqn:Planck_5:phi}
\ea
This implies that
\ba
\left|{\mu _{eff} ^2\over {M_{\phi}\lambda _{eff} }}\right|
\sim {1\over {\xi}}M_{Pl(4)} ^2 \gg TeV ^2,
\label{eqn:Planck_4}
\ea
which renders the brane mediated mechanism of SSB
unviable for phenomenological reasons.
This means that the phenomenological hierarchy between the SM
typical energy scale of order $TeV$ and the Planck scale of the
gravitational effects of the physics on the brane cannot be accounted
by the SSB brane mechanism, since the characteristic scale
of the induced dynamics of the scalar field is the Planck scale.
It is easy to see that $\left< \Phi\right> \sim M_{Pl(4)}.$ Thus, the
scalar field becomes a short range
field about the brane and therefore strongly localized therein.
However, it is the non-minimal coupling that, upon spontaneous
symmetry breaking, allows the matter localized on the
brane to interact with bulk matter fields with typically gravitational
strength.

Moreover, from the expression for the five-dimensional cosmological
constant in the case of a vanishing effective cosmological constant,
we find that $\Lambda _{(5)} \sim -M_{\phi}V ( \left< \Phi \right> ^2)
-\sigma ^2 \kappa _{(5)}^2,$
and consequently that
\ba
\sigma ^2 \sim M_{Pl(4)} ^2 M_{\phi}\left[
-\Lambda _{(5)} -M_{\phi}V ( \left< \Phi \right> ^2)\right] ~.
\label{eqn:Planck_5:Lambda_5}
\ea

\subsection{The effective potential as a measure of the
  thickness of the brane}

In what follows we argue that the mass scale generated by the SSB mechanism
sets the range of the fields on the brane and hence the thickness of
the brane. In order to understand the role of the mass  $M_{\phi},$ we
expand the vev of $\Phi$ observed on the brane about the value expected in its
absence. Defining
\ba \mu
_{\sigma}^2 =2\sigma ^2\xi \kappa _{(5)}^4,
\quad \lambda _{\sigma} =M_{\phi}\sigma ^2\xi
^2\kappa _{(5)}^6,
\ea
then equations (\ref{eqn:mu:eff2}) and (\ref{eqn:lambda:eff2}) become
\ba
\mu _{eff}^2 &\sim& \mu_{(5)}^2 +\mu _{\sigma}^2,\\
M_{\phi}\lambda _{eff} &\sim& M_{\phi}\lambda _{(5)} +\lambda _{\sigma}.
\ea
It follows that
\ba
-\left< \Phi \right>^2
={\mu _{eff} ^2\over {M_{\phi}\lambda _{eff}}}
&\sim& (\mu _{(5)}^2 + \mu _{\sigma}^2){1 \over {M_{\phi}\lambda _{(5)}}}
 \left( 1 -{\lambda _{\sigma}\over {M_{\phi}\lambda _{(5)}}}
  +O\left[ \left(
    {\lambda _{\sigma}\over {M_{\phi}\lambda _{(5)}}}\right)^2\right]
  \right)\cr
&\sim& {\mu _{(5)}^2\over {M_{\phi}\lambda _{(5)}}}
-{\mu _{\sigma}^2\over {M_{\phi}\lambda _{(5)}}}
  {\lambda _{\sigma}\over{M_{\phi}\lambda _{(5)}}}
+{\mu _{\sigma} ^2 \over {M_{\phi}\lambda _{(5)}}}
+O\left[ ( \mu _{\sigma}^2)^2, \lambda _{\sigma} ^2\right]
\ea
keeping only the lowest power in the parameters indexed $\sigma.$
We can now distinguish between two cases, depending on which
$\sigma$-indexed term on the right hand side dominates.

For the case when the second term dominates over the third, then
\ba
{\mu _{eff}^2 \over {M_{\phi}\lambda _{eff}}}
> {\mu _{(5)}^2\over {M_{\phi}\lambda _{(5)}}}
\Rightarrow \left| {\mu _{eff}^2 \over {M_{\phi}\lambda _{eff}}} \right|
<\left| {\mu _{(5)}^2\over {M_{\phi}\lambda _{(5)}}}\right|.
\ea
For an increasingly important
contribution of the brane tension, both the minimum and the inflexion point of
the effective potential converge to $\left <\Phi \right> \to 0,$ whereas
$V_{eff} \to \lambda _{eff}(\Phi^4/4).$

For the case when the third term dominates over the second, then the opposite
relation holds and the minimum and the inflexion points become larger with an
increasing brane contribution.
We observe that the presence of the brane affects the characteristics of the
potential and consequently the SSB mechanism and the scales generated thereby.
Thus, it is in order to compare 
$\left <\Phi \right>$ with the inverse of the thickness of the brane,
estimated to be of order the effective mass $\mu _{eff}$. 
We find that $M_{\phi}\lambda _{eff} \sim 1.$ From Eq.~(\ref{eqn:Planck_4}),
this also implies that $\mu _{eff} ^2 \sim M_{Pl(4)}^2.$

The present analysis can be extended to a complex scalar field $\Psi,$
coupled to a $U(1)$ gauge, resulting quite similar conclusions \cite{BC07}.
Both the complex scalar field and the $U(1)$ vector field acquire, upon SSB
and through a non-minimal coupling of the form $\Psi \bar{\Psi} R,$ masses of
order the Planck mass as well.

\section{Discussion}

In this contribution we have examined the mechanism of spontaneous symmetry
breaking due to a scalar field in the bulk spacetime coupled non-minimally
to gravity.
We have shown that a non-minimal coupling can be a source
of symmetry breaking on the brane but only at very high energies.
We derived the conditions which allow for the existence of a
non-vanishing bulk scalar field vacuum configuration
and demonstrated that the scales of the induced masses are of order the
four-dimensional Planck scale, thus failing to accommodate on the
brane the typical scales
of the Standard Model.
We notice, however, that this implies that the bulk scalar fields become very
short range about the position of the brane and thus strongly localized
therein.

Furthermore, we observe that in the absence of the non-minimal coupling of the
bulk scalar  fields to gravity, i.e. for $\xi =0,$  the effective potential on
the brane of a bulk scalar field  reduces in both cases to  
$V_{eff} = \mu _{(5)} ^2( \phi ^2/2) +\lambda _{(5)}( \phi ^4/4) 
+(3/4)\sigma ^2 \kappa _{(5)}^2d^2/[(d +1)(d -1)].$  
The realization of a braneworld universe  as a
codimension-one surface of localized matter contributes a constant term
proportional to the brane tension $\sigma$ to the effective potential.  The
brane tension does not, however, contribute to the mechanism of spontaneous
symmetry breaking observed on the brane unless the bulk scalar fields are
non-minimally coupled to gravity. This is observed in the dependence on the
coupling parameter $\xi$ of the parameters $\mu _{eff}^2$ and $\lambda_{eff}.$
Moreover, the mixing of the discontinuity in  the extrinsic curvature with the
discontinuity in the normal derivative of the scalar field, as encompassed by
the corresponding  matching conditions, is also $\xi$--dependent.  Such mixing
is switched off when  $\xi =0,$ as  already noticed in Ref.~{\cite{BV00}} and
also found in Ref.~{\cite{BC06}}. These observations seem to suggest that the
matter localized on the brane will interact with bulk matter fields through
gravity only if a non-minimal coupling exists. We have also argued that
the characteristics of the potential, and in particular the vev of the scalar
field, can be related to the estimated thickness of the brane,
given that the former sets the effective range of the fields. This suggests
that the SSB mechanism is closely related to the localization process of
fields on the brane.

On the technical side, our approach goes a step further in setting up the
framework for studying the brane induced physics which arises from the
presence of matter fields in the bulk. The case of a vector field coupled
non-minimally to gravity was previously considered and its implications for
Lorentz symmetry on the brane studied in Ref.  \cite{BC06}. The case of the
scalar field, both real and complex, coupled non-minimally to gravity is
treated in Ref. \cite{BC07}. Our approach allows to relate the cosmological
constant problem and the scale of gravity to the mechanism of origin of
mass, which suggests to relate to the process of localization of bulk
fields on the brane by setting the nature of the interaction between brane and
bulk fields to be essentially gravitational.

\vskip 0.5cm

\centerline{\bf {Acknowledgments}}

\vskip 0.3cm

\noindent C. C. thanks Funda\c c\~ao para a Ci\^encia e a Tecnologia - FCT
(Portuguese Agency) for financial support under the fellowship
/BPD/18236/2004. O. B. acknowledges the partial support of the FCT project
POCI/FP/63916/2005 and of the European Science Foundation network programme
"Quantum Geometry and Quantum Gravity". O. B. also expresses his gratitude to 
Florian Girelli and 
Stefano Liberati for the kind hospitality at Trieste. 
The authors thank Martin
Bucher, Mariam Bouhmadi Lopez, Fernando Quevedo and Kyriakos Tamvakis for
useful discussions.



\appendix
\section{Geometry and Matter Sources}

In this section we systematize the procedure that we used to derive the
equations of motion on a codimension-one brane from a given bulk action.

We begin to parametrize the worldsheet in terms of coordinates
$x^{A}=(t_b,{\bf x}_b)$ intrinsic to the brane \cite{bucher05}. Using
the chain rule, we may express the brane tangent and normal unit
vectors in terms of the bulk basis as follows:
\begin{eqnarray}
\hat e_A &=&{\partial \over \partial x^A}
=X_A^\mu{\partial \over \partial x^\mu} =X_A^\mu \hat e_\mu,\cr
\hat e_N &=&{\partial \over \partial n}
=N^\mu {\partial \over \partial x^\mu} =N^\mu \hat e_\mu,
\end{eqnarray}
with
\begin{eqnarray}
g_{\mu\nu}N^\mu N^\nu =1,\quad g_{\mu\nu}N^\mu X^\nu_A =0,
\end{eqnarray}
where ${\bf g}$ is the bulk metric
\begin{eqnarray}
{\bf g}
&=&g_{\mu\nu}{\hat e}_\mu\otimes {\hat e}_\nu
=g_{AB}~{\hat e}_A\otimes {\hat e}_B
+g_{AN}~{\hat e}_A\otimes {\hat e}_N
+g_{NB}~{\hat e}_N\otimes {\hat e}_B
+g_{NN}~{\hat e}_N\otimes {\hat e}_N.
\end{eqnarray}
To obtain the metric induced on the brane we expand the bulk basis
vectors in terms of the coordinates intrinsic to the brane and keep the
doubly brane tangent components only. It follows that
\begin{eqnarray}
g_{AB} =X^\mu _AX^\nu _B~g_{\mu\nu}
\end{eqnarray}
is the $(3+1)$-dimensional
metric induced on the brane by the $(4+1)$-dimensional bulk metric
$g_{\mu\nu}.$
The induced metric with upper indices is defined by the relation
\begin{eqnarray}
g_{AB}~g^{BC}=\delta _A{}^C.
\end{eqnarray}
It follows that we can write any bulk tensor field as a linear
combination of mutually orthogonal vectors on the brane, $\hat e_A,$
and a vector normal to the brane, $\hat e_N.$ We illustrate the
example of a vector $B_{\mu}$ and a tensor $T_{\mu\nu}$ bulk fields as follows
\begin{eqnarray}
{\bf B}
&=& B_{A}~{\hat e}_A +B_{N}~{\hat e}_N, \\
{\bf T}
&=&T_{AB}~{\hat e}_A\otimes {\hat e}_B
+T_{AN}~{\hat e}_A\otimes {\hat e}_N
+T_{NB}~{\hat e}_N\otimes {\hat e}_B
+T_{NN}~{\hat e}_N\otimes {\hat e}_N.
\end{eqnarray}
Derivative operators decompose similarly.
We write the derivative operator $\nabla$ as
\begin{eqnarray}
\nabla =(X^\mu_A +N^\mu)\nabla_\mu =\nabla_A +\nabla_N.
\end{eqnarray}
Choosing Fermi normal coordinates for the neighborhood of a point on
the brane, all the Christoffel symbols of the metric on the boundary are
zero. The partial derivatives, however, do not vanish in general.
The continuation of the coordinates off the boundary is given by
the Gaussian normal prescription.
The non-vanishing connection coefficients are thus
\begin{eqnarray}
\nabla_A {\hat e}_B &=&-K_{AB}~{\hat e}_N, \cr
\nabla_A {\hat e}_N &=&+K_{AB}~{\hat e}_B, \cr
\nabla_N {\hat e}_A &=&+K_{AB}~{\hat e}_B, \cr
\nabla_N {\hat e}_N &=&0.
\end{eqnarray}
For the derivative operator $\nabla\nabla$ we find that
\begin{eqnarray}
\nabla\nabla&=& g^{\mu\nu}\nabla_\mu \nabla_\nu \cr
&=&g^{AB}\left[ (X^\mu_A\nabla_\mu)(X^\nu_B\nabla_\nu)
-X^\mu_A(\nabla_\mu X^\nu_B)\nabla_\nu \right]
+g^{NN}\left[ (N^\mu\nabla_\mu)(N^\nu\nabla_\nu)
-N^\mu(\nabla_\mu N^\nu)\nabla_\nu \right] \cr
&=&g^{AB}\left[ \nabla_A \nabla_B +K_{AB}\nabla_N\right]
+               \nabla_N \nabla_N.
\end{eqnarray}
We can now decompose the Riemann tensor, $R_{\mu\nu\rho\sigma},$ along
the tangent and the normal directions to the surface of the brane as
follows
\begin{eqnarray}
R_{ABCD} &=& R_{ABCD}^{(ind)} +K_{AD}K_{BC} -K_{AC}K_{BD}, \qquad\\
R_{NBCD} &=& K_{BC;D} -K_{BD;C}, \\
R_{NBND} &=& K_{BC}K_{DC} -K_{BD,N},
\end{eqnarray}
from which we find the decomposition of the Einstein tensor,
$G_{\mu\nu},$ obtaining the Gauss-Codacci relations
\begin{eqnarray}
G_{AB} &=&G_{AB}^{(ind)} +2K_{AC}K_{B}{}^{C} -K_{AB}K -K_{AB,N}
-{1\over 2}g_{AB}\left( -K_{CD}K^{CD} -K^2 -2K_{,N}\right),\quad \\
G_{AN} &=&K_{A}{}^{B}{}_{;B} -K_{;A}, \\
G_{NN} &=&{1\over 2}\left( -R^{(ind)} -K_{CD}K^{DC} +K^2\right).
\label{eqn:gc}
\end{eqnarray}
We have now the tools to express the bulk equations of motion derived
from the bulk action 
as a decomposition along the parallel and orthogonal directions to the
brane as defined by 
the Gaussian normal prescription.

Next we define the boundary conditions compatible with the presence of the
brane.
We regard the brane as shell of thickness $2\delta$ in the limit
$\delta \to 0$ and separating the bulk into two mirroring regions. 
Two consequences follow.
Derivatives of quantities discontinuous across the brane generate
singular distributions on the brane. The integration of these terms in the
coordinate normal to the brane relates the induced geometry
with the localization of the induced stress-energy
in the form of matching conditions.
Moreover, the $Z_2$-symmetry establishes the continuity conditions for
the fields across the brane.
From the continuity of the quantities on the brane,
it follows that the parallel components must be even in $N$ and
consequently that the orthogonal components must be odd in $N,$ with
each additional orthogonal component reverting the parity.
[See Ref.~\cite{gherghetta00} for odd fields about the brane.]
In particular, for a scalar field $\phi$ and a vector field ${\bf B}$
we have that
\ba
\phi (-\delta) =&+\phi (+\delta),\qquad
  (\nabla_{N}\phi)(-\delta) &=-(\nabla_{N}\phi)(+\delta),\cr
B_{A}(-\delta) =&+B_{A}(+\delta),\qquad
  (\nabla_{N}B_{A})(-\delta) &=-(\nabla_{N}B_{A})(+\delta),\cr
B_{N}(-\delta) =&-B_{N}(+\delta),\qquad
  (\nabla_{N}B_{N})(-\delta) &=+(\nabla_{N}B_{N})(+\delta).
\ea
Similarly, we
have for the metric that
$g_{AB}(N=-\delta) =+g_{AB}(N=+\delta),$
which implies that
\ba
K_{AB}(N=-\delta) =-K_{AB}(N=+\delta).
\ea

After extracting the singular terms from the projected equations and
into the matching  condition, we obtain the equations of motion
induced on the brane. Since the original  equations were derived from
a bulk action, the induced ones will be expressed in terms of  bulk
parameters, which must be rescaled in order to reproduce the observed
characteristically four-dimensional parameters. The rescaling
parameter is expected to be related to a measure of the thickness of
the brane.

\end{document}